\documentclass[aps,prl,twocolumn,superscriptaddress,showpacs,floatfix,letterpaper]{revtex4-2}%

\usepackage{graphicx}
\usepackage{tabularx}
\usepackage{bm}
\usepackage{latexsym}
\usepackage{epsf}
\usepackage{rotating}
\usepackage{epsfig,graphics,rotate,color}
\usepackage{wrapfig}
\usepackage{amssymb}
\usepackage{amsmath}
\usepackage{amsfonts}
\usepackage{array,hhline,dcolumn}%
\usepackage[normalem]{ulem}
\usepackage{color}
\usepackage[colorlinks=true,allcolors=blue]{hyperref}

\begin{document}
\bibliographystyle{apsrev4-1}

\title{Neutrino Decoherence and the Mass Hierarchy in the JUNO Experiment}






\author{E. Marzec}
\affiliation{
University of Michigan, Ann Arbor, MI 48109, USA
}

\author{J. Spitz}
\affiliation{
University of Michigan, Ann Arbor, MI 48109, USA
}

\begin{abstract}
The finite size of a neutrino wavepacket at creation can affect its oscillation probability. Here, we consider the electron antineutrino wavepacket and decoherence in the context of the nuclear reactor based experiment JUNO. Given JUNO's high expected statistics [$\sim$100k IBD events ($\overline{\nu}_e p \rightarrow e^+ n$)], long baseline ($\sim$53\,km), and excellent energy resolution [$\sim$$0.03/\sqrt{E_{\mathrm{vis}}~\mathrm{(MeV)}}$], its sensitivity to the size of the wavepacket is expected to be quite strong. Unfortunately, this sensitivity may weaken the experiment's ability to measure the orientation of the neutrino mass hierarchy for currently allowed values of the wavepacket size. Here, we report both the JUNO experiment's ability to determine the hierarchy orientation in the presence of a finite wavepacket and its simultaneous sensitivity to size of the wavepacket and the hierarchy. We find that wavepacket effects are relevant for the hierarchy determination up to nearly two orders of magnitude above the current experimental lower limit on the size, noting that there is no theoretical consensus on the expectation of this value. We also consider the effect in the context of other aspects of JUNO's nominal three-neutrino oscillation measurement physics program and the prospect of future enhancements to sensitivity, including from precise measurements of $\Delta m^2_{3l}$ and a near detector.
\end{abstract}
\maketitle

\section{Introduction}
The plane-wave treatment of neutrino oscillations, in which the propagating neutrino is assigned a definitive momentum, is an excellent approximation for terrestial- and atmospheric-based neutrino experiments. This framework breaks down, however, in the case that the neutrino wavepacket is finite (see, e.g. Refs.~\cite{Kiers:1995zj,PhysRevD.24.110,Giunti:1997wq,Giunti:2003ax}). The different velocities of the propagating mass eigenstates leads to their separation and changes the oscillation probability, with the effect increasing for larger travel distances, lower energies, and larger mass splittings. All observed neutrino oscillation signatures, with the possible exception of solar and supernova neutrinos which are expected to be completely decoherent, are so far consistent with coherent neutrino oscillations due to the relatively small size of the observable contributions compared to experimental resolutions. However, the increasing capability of neutrino experiments makes gaining sensitivity to the finite size of the wavepacket a reasonable future possibility.

Originating inside of a nuclear reactor and with contributions from the beta decays of some $\sim$1000 isotopes~\cite{Huber:2011wv}, the typical size of an electron antineutrino wavepacket at creation, denoted by $\sigma_x$, is at present unknown. However, there is potential for performing this detailed and complex calculation in the future, ala Ref.~\cite{Jones:2014sfa}. For now, the distance scales of the decay, including the characteristic beta-decay-nucleus size ($\sim$$10^{-5}$\,nm) and the inverse of the antineutrino energy ($\sim$$10^{-4}$\,nm)~\cite{deGouvea:2021uvg}, considered alongside the current reactor-based limits ($\sigma_x>2.1\cdot 10^{-4}$~nm at 90\%~CL using a phenomenological combination of Daya Bay, KamLAND, and RENO data~\cite{deGouvea:2021uvg}, and $\sigma_x>1\cdot 10^{-4}$\,nm at 95\% CL from a dedicated measurement with Daya Bay~\cite{DayaBay:2016ouy}), motivate a study of the effects of $\sigma_x$ on experimental observables for values as low as $10^{-4}$\,nm. We note, however, that the other relevant distance scale in the decay, the inter-atomic spacing of the Uranium-based fuel ($\sim$0.1-1\,nm), is well above the level at which wavepacket effects would be discernible in any realistic future reactor-based experiment. 

Expecting first data in 2023, the ambitious JUNO project is at the forefront of what is possible with a reactor antineutrino experiment~\cite{JUNO:2015zny,JUNO_2022_physics_paper}. The experiment will feature a 20\,kton liquid scintillator far detector about 53\,km from a 26.6\,GWth total set of PWR reactor complexes. With 77.9\% photocoverage, JUNO can expect $\sim$$0.03/\sqrt{E_{\mathrm{vis}}~\mathrm{(MeV)}}$ positron visible energy resolution for characterizing the $\sim$100k IBD events ($\overline{\nu}_e p \rightarrow e^+ n$) expected in the data collection period (6~years). As has been shown in Refs.~\cite{deGouvea:2020hfl,JUNO:2021ydg}, the high statistics, long baseline, and strong energy resolution of JUNO translate to more than an order of magnitude better sensitivity to the wave packet effect as compared to previous reactor experiments, including Daya Bay~\cite{dayabay}, KamLAND~\cite{KamLAND:2008dgz}, and RENO~\cite{reno}. 

The effect of a finite wavepacket size on the oscillation probability is easy to demonstrate by considering atmospheric-only electron-flavor disappearance (with normal mass ordering):
\begin{equation}
 P_{ee}\rightarrow 1- \frac{1}{2}\sin^2 2\theta_{13}\Big[1-\cos\frac{1.27\Delta m^2_{31}L}{E}\mathrm{exp}(-\frac{L^2(\Delta m^2_{31})^2}{32E^4\sigma^2_{x}})\Big] \label{simple_osc} 
\end{equation}

We see that the dampening of the oscillation probability due to the wavepacket size ($\sigma_x$) increases with $L$, decreases with $E$, and increases with $\Delta m^2$. As such, the effect is larger for sterile ($\Delta m^2\sim 1~\mathrm{eV}^2$)~\cite{Arguelles:2022bvt} and atmospheric oscillations ($\Delta m^2\sim 3\cdot 10^{-3}~\mathrm{eV}^2$) as compared to solar oscillations ($\Delta m^2\sim 7\cdot 10^{-5}~\mathrm{eV}^2$). JUNO's large $L$ ($\sim$53\,km), low $E_{\overline{\nu}_e}$ ($\sim$2-10\,MeV), and sensitivity to the atmospheric mass splitting makes it an excellent testing ground for the effect of $\sigma_x$ on oscillation behavior. In general, one can expect JUNO to be more sensitive to wavepacket effects within the three neutrino oscillation paradigm~\cite{Arguelles:2022bvt}, at least, than any existing or even near-term-planned experiment.

In addition to the wavepacket effect, JUNO's extraordinary capability also makes it sensitive to the orientation of the neutrino mass hierarchy, among other oscillation parameters~\cite{JUNO:2015zny,JUNO_2022_physics_paper}. The three-neutrino oscillation probability in vacuum and without the wavepacket effect is given by $P_{ee}=1-P_{21}-P_{31}-P_{32}$, with
\begin{eqnarray}
P_{21}&& =\cos^4\theta_{13}\sin^2 2\theta_{12}\sin^2\Big(\frac{1.27\Delta m^2_{21}L}{E}\Big) \nonumber\\ \nonumber \\
P_{31}&& =\cos^2\theta_{12}\sin^2 2\theta_{13}\sin^2\Big(\frac{1.27\Delta m^2_{31}L}{E}\Big) \nonumber\\ \nonumber \\
P_{32}&& =\sin^2\theta_{12}\sin^2 2\theta_{13}\sin^2\Big(\frac{1.27\Delta m^2_{32}L}{E}\Big)~.  
\label{Pee32}
\end{eqnarray} 

\begin{figure*}[ht]
\begin{center}
\includegraphics[width=0.49\textwidth]{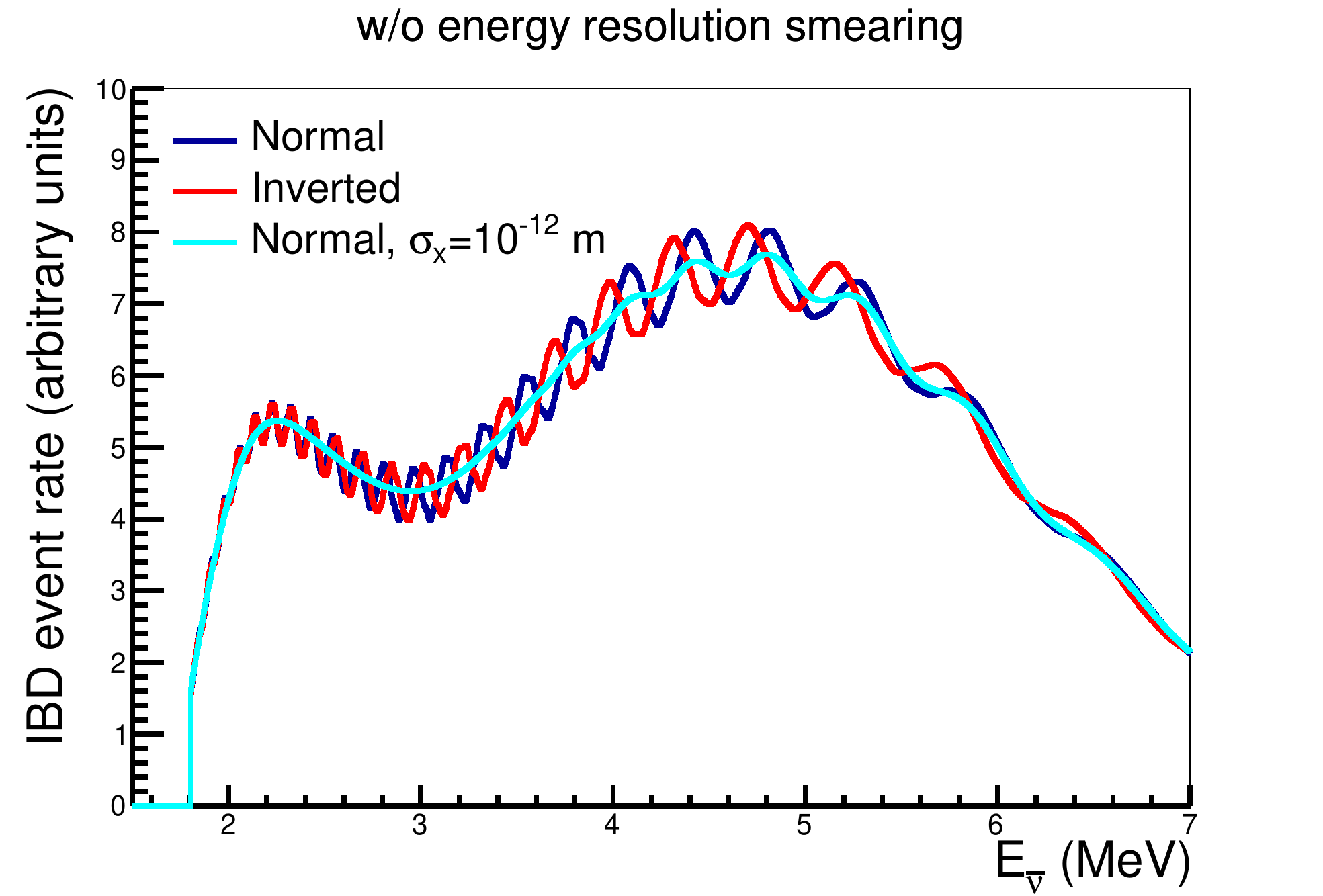}
\includegraphics[width=0.49\textwidth]{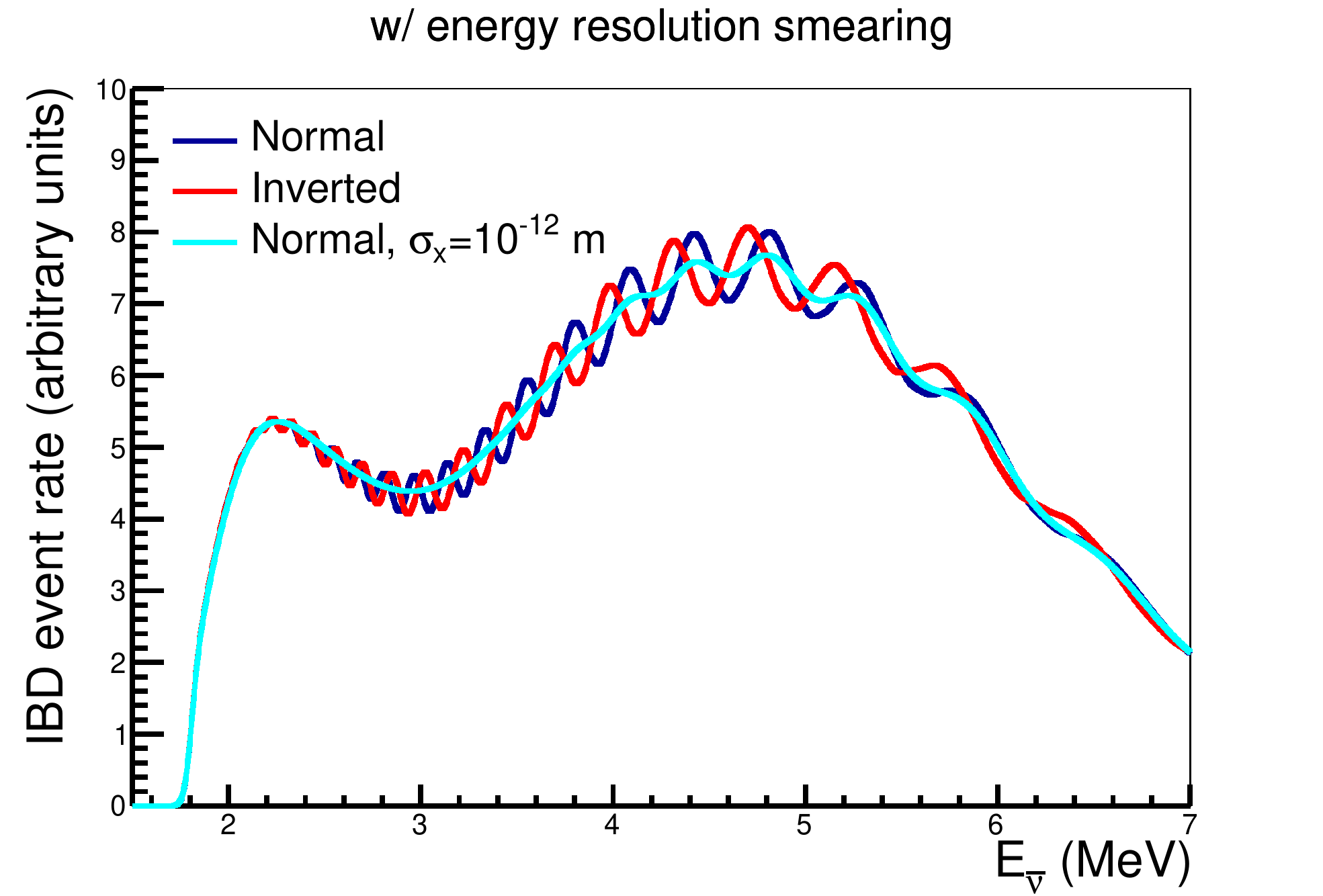}
\end{center}
\vspace{-.5cm}
\caption{The IBD event rate distribution shapes expected at JUNO, without (left) and with (right) energy smearing due to reconstruction resolution. The shape differences between the normal and inverted hierarchy are shown. In addition, the dampening effect of finite $\sigma_x$ ($10^{-12}$\,m) on the normal hierarchy oscillation probability can be seen.}
\label{rate_plot}
\end{figure*}

The difference in oscillation probability between the normal hierarchy ($m_3>m_2>m_1$) and inverted hierarchy ($m_2>m_1>m_3$) for the JUNO experiment can be seen in Figure~\ref{rate_plot} in terms of IBD-based reconstructed antineutrino energy ($E_{\bar{\nu}_e}=E_{e^+}+0.78$\,MeV), with and without energy resolution smearing according to the expected positron energy resolution (see below).

The discernible shape difference between the hierarchies is reduced with oscillation dampening due to a finite wavepacket size. The three-neutrino oscillation equation (Eq.~\ref{Pee32}) is modified to include wavepacket effects with
\begin{eqnarray}
P_{21}&& =\cos^4\theta_{13}\sin^2 2\theta_{12}\cdot \frac{1}{2}\Big[1-\cos\frac{1.27\Delta m^2_{21}L}{E}\cdot\nonumber\\&&\mathrm{exp}(-\frac{L^2(\Delta m^2_{21})^2}{32E^4\sigma^2_{x}})\Big]\nonumber \\
P_{31}&& =\cos^2\theta_{12}\sin^2 2\theta_{13}\cdot \frac{1}{2}\Big[1-\cos\frac{1.27\Delta m^2_{31}L}{E}\cdot\nonumber\\&&\mathrm{exp}(-\frac{L^2(\Delta m^2_{31})^2}{32E^4\sigma^2_{x}})\Big] \nonumber \\
P_{32}&& =\sin^2\theta_{12}\sin^2 2\theta_{13}\cdot \frac{1}{2}\Big[1-\cos\frac{1.27\Delta m^2_{32}L}{E}\cdot\nonumber\\&&\mathrm{exp}(-\frac{L^2(\Delta m^2_{32})^2}{32E^4\sigma^2_{x}})\Big] ~.
\label{Pee32_sigmax}
\end{eqnarray} 

Figure~\ref{rate_plot} shows the dampening effect of $\sigma_x=10^{-12}$\,m, for example, on the oscillation probability in the case of a normal hierarchy. As can be seen, the fast-atmospheric oscillation probability, with frequency governed by $\Delta m^2_{3l}$~\footnote{We adopt the NuFIT notation here: $l=1$ corresponds to $\Delta m^2_{3l}>0$ (normal hierarchy) and $l=2$ corresponds to $\Delta m^2_{3l}<0$ (inverted hierarchy)}, is washed out and the power to distinguish the hierarchies is significantly reduced, especially at lower antineutrino energies where the effect is more pronounced. Fortunately, however, while the amplitude of oscillation decreases, the oscillation frequency and phase is left unchanged by the wavepacket size. Therefore, while the finite size of the wavepacket can lead to diminished, or even nullified (in the case that the fast atmospheric oscillations completely disappear), sensitivity to a hierarchy determination, it is unlikely to lead to a confused or systematically incorrect determination. 

Previous work has considered the effect of the wavepacket size on hierarchy determination in generic medium baseline reactor experiments~\cite{Chan:2015mca}, and JUNO's sensitivity to the size of the wavepacket with a known hierarchy~\cite{deGouvea:2020hfl,JUNO:2021ydg}. Below, we report the JUNO experiment's ability to determine the hierarchy in the presence of a finite wavepacket, with an emphasis on \textit{experimentally allowed} values of this parameter, again noting that there is no reliable theoretical prediction, and consider JUNO's sensitivity to both the wavepacket size and the orientation of the hierarchy at the same time. The experimental assumptions and analysis methods are detailed in the next section, followed by a section presenting the results and discussion, and then conclusions.

\section{Analysis}
\label{analysis}

To model the JUNO experiment, we use the reactor electron antineutrino spectrum
as described in Ref.~\cite{Vogel:1989iv} with a PWR fuel mixture consistent with Ref.~\cite{JUNO:2015zny} ($\mathrm{^{235}U}:\mathrm{^{238}U}:\mathrm{^{239}Pu}:\mathrm{^{241}Pu}=0.577:0.076:0.295:0.052$). Nine relevant PWR nuclear reactor complexes create the antineutrino flux, and their powers and distances from the JUNO detector are modeled according to Ref.~\cite{JUNO_2022_physics_paper}. The largest contributions come from the Yangjiang (61.5\% of total flux) and Taishan (32.1\%) Nuclear Power Plant complexes at around a 53~km baseline.  Uncertainties from the reactor flux shape are expected to contribute at the sub-1\% level with the inclusion of data from a near detector called ``JUNO-TAO"~\cite{JUNO:2020ijm}. Construction of JUNO-TAO is ongoing and first data is expected in 2022. The existence of this near detector in constraining the flux shape uncertainty is implicitly assumed throughout this analysis. At the sub-1\% level, the reactor flux shape uncertainties are not expected to significantly affect either the hierarchy determination or sensitivity to $\sigma_x$. Similarly, the reactor flux normalization uncertainty contribution to the sensitivity can be considered negligible~\cite{JUNO:2015zny}. We ignore both sources of flux uncertainty in this analysis.

After creation, the electron antineutrinos are propagated from each reactor, appropriately weighted by distance and power, to the JUNO far detector while applying the three-neutrino oscillation equation, including the different baselines (Eq.~\ref{Pee32_sigmax}). The central-value three-neutrino oscillation parameters and uncertainties are taken from NuFIT\,5.0, including Super-K atmospheric data~\cite{Esteban:2020cvm}. Correlations among the parameters are ignored when sampling from the allowed ranges. Also, while terrestrial matter effects provide corrections to $\sin^2 \theta_{12}$ and $\Delta m^2_{21}$ at the 0.5-1.0\% level for these distances and energies, they are negligible for $\sin^2 \theta_{13}$ and $\Delta m^2_{3l}$, and the associated hierarchy and $\sigma_x$ discussion, and are ignored here for simplicity.

In addition to the standard oscillations, $\sigma_x$ values from $10^{-13}$-$10^{-9}$\,m are considered as contributing to the effective oscillation probability. The lower end of this range roughly corresponds to the current experimental limit ($\sigma_x>2.1\cdot 10^{-4}$~nm at 90\% CL~\cite{deGouvea:2021uvg}) and the upper end roughly corresponds to the largest relevant distance scale of the antineutrino-producing beta decay(s), the upper end of the inter-atomic spacing of the Uranium-based fuel. We consider a single $\sigma_x$ value as affecting the oscillation probability of all $\overline{\nu}_e$ from all reactor complexes considered here noting that a, likely small, range of relevant $\sigma_x$ values will contribute in reality given the different environmental conditions affecting the size of the wavepacket in each reactor core and even individual isotope. For simulating electron antineutrino induced IBD events, we use the IBD cross-section from Ref.~\cite{Vogel:1999zy}. The resulting positron energy spectrum is smeared by the following energy resolution to simulate JUNO's energy reconstruction capability~\cite{JUNO_2022_physics_paper}:
\begin{equation}
\frac{\sigma_{E_{\mathrm{vis}}}}{E_{\mathrm{vis}}}=\sqrt{\Big(\frac{a}{\sqrt{E_{\mathrm{vis}}}}\Big)^2+b^2+\Big(\frac{c}{E_{\mathrm{vis}}}\Big)^2}~,
\label{resolution}
\end{equation}
with $a=2.61\%,~b=0.82\%,~c=1.23\%$ and $E_{\mathrm{vis}}$ in MeV. Notably, the non-linear response of the liquid scintillator due to the ``quenching effect" is not accounted for in this resolution estimate. The reader is referred to Refs.~\cite{JUNO_2022_physics_paper,DayaBay:2019fje} for more on this relevant issue, which is ignored here for simplicity.  

\begin{table} [t]
\begin{center}
\begin{tabular}{|c|c|}
\hline
Experimental parameters &   \\ \hline
Total IBD events ($\sim$6 years) &  $10^5$ (post-oscillations)  \\ 
IBD cross section shape &  Vogel \& Beacom~\cite{Vogel:1999zy}  \\ 
$e^+$ visible energy resolution &See Eq.~\ref{resolution} \\  
Near detector?  &  Yes, JUNO-TAO~\cite{JUNO:2020ijm}  \\  \hline
Reactor flux shape  &  Vogel \& Engel~\cite{Vogel:1989iv}  \\
Fuel composition  &    \\
$\mathrm{^{235}U}:\mathrm{^{238}U}:\mathrm{^{239}Pu}:\mathrm{^{241}Pu}$  &   $0.577:0.076:0.295:0.052$   \\ 
\hline
9 reactor complexes  &  [4.6, 52.77], [4.6, 52.64],    \\
{[}power (GWth), baseline (km){]} &  [2.9, 52.74], [2.9, 52.82],   \\
  &  [2.9, 52.41], [2.9, 52.49],   \\
    &  [2.9, 52.11], [2.9, 52.19],   \\
        &  [17.4, 215]   \\\hline \hline
Oscillation parameters (3$\nu$) &   \\ \hline
$\theta_{12}$ ($^\circ$)&  $33.44^{+0.77}_{-0.74}$  \\ 
$\theta_{23}$ ($^\circ$)&  $49.2^{+1.0}_{-1.3}$  \\
$\theta_{13}$ ($^\circ$)&  $8.57^{+0.13}_{-0.12}$  \\ 
$\Delta m^2_{21}$ ($\mathrm{eV}^2$)&  $(7.42^{+0.21}_{-0.20})\cdot 10^{-5}$  \\ 
$\Delta m^2_{31_\mathrm{NH}}$ ($\mathrm{eV}^2$)&  $(2.515\pm0.028)\cdot 10^{-3}$  \\ 
$\Delta m^2_{32_\mathrm{IH}}$ ($\mathrm{eV}^2$)&  $(-2.498^{+0.028}_{-0.029})\cdot 10^{-3}$  \\  \hline
\end{tabular} \caption{Summary of the relevant experimental parameter assumptions.}\label{table:values}
\end{center}
\end{table}

The representative JUNO dataset size is assumed to be 100k electron antineutrino IBD events after standard oscillations~\cite{JUNO_2022_physics_paper}, from 6~years of running, and we consider this sample in 200~equally-spaced reconstructed antineutrino energy bins from 1.8 to 8\,MeV. The statistical and shape uncertainties from all backgrounds (mainly originating from geo-neutrinos, fast neutrons, accidentals, and the decays of cosmogenic-induced isotopes) are expected to have a minimal impact on hierarchy sensitivity~\cite{JUNO:2015zny} and are ignored here. A summary of the experimental parameter assumptions is shown in Table~\ref{table:values}.

With the goal of differentiating one hierarchy from another, we follow Ref.~\cite{JUNO:2015zny} in forming a $\chi^2$ test statistic, comparing the simulated ``observed" events ($O$) in each reconstructed electron antineutrino energy bin ($i$) to a ``prediction" ($P$) under a particular oscillation+wavepacket scenario:
\begin{equation}
\chi^2=\sum^{200}_{i=1}  \frac{[O_i-P_i(1+\sum_k \alpha_{ik}\epsilon_{k})]^2}{P_i}+\sum_{k}\frac{\epsilon_{k}^2}{\delta_{k}^2}~. 
\end{equation}

A ``fake data set" observed spectrum ($O_i$) is created with a particular hierarchy and $\sigma_x$ value, along with oscillation parameters sampled according to their central values and uncertainties shown in Table~\ref{table:values}, and the statistics expected in the experiment. The prediction ($P_i$) is varied to find the $\chi^2_{min}$, while pull terms on the oscillation parameters ($\epsilon_k$, with an $\alpha_{ik}$ fractional contribution to $P_i$) and associated uncertainties ($\delta_k$) (shown in Table~\ref{table:values}) are used to constrain the prediction in the minimization procedure, and the $\sigma_x$ parameter is left unconstrained. 

Sensitivity to the mass hierarchy determination can be quantified by comparing the $\chi^2_{\mathrm{min}}$ with a normal hierarchy based prediction to the $\chi^2_{\mathrm{min}}$ with an inverted hierarchy based prediction:
\begin{equation}
\Delta\chi^{2}_{\mathrm{MH}}=\chi^2_{\mathrm{min}}(\mathrm{NH})-\chi^2_{\mathrm{min}}(\mathrm{IH})
\end{equation}

$\Delta\chi^{2}_{\mathrm{MH}}$ will strongly depend on whether the observed spectrum ($O_i$) was produced with an underlying normal hierarchy or an underlying inverted hierarchy. That is, the sensitivity to hierarchy determination can be dependent on the true underlying hierarchy. We use a subscript on $\Delta \chi^2$ to denote the underlying ``true" orientation used to produce the observed sample (i.e. either $\Delta \chi^2_{\mathrm{NH}}$ or $\Delta \chi^2_{\mathrm{IH}}$).

To summarize, we test JUNO's ability to resolve the mass hierarchy for an unknown 
$\sigma_{x}$ by considering many different underlying ``true" $\sigma_{x}$ values with a known hierarchy. At each scanned true value of $\sigma_{x}$, a JUNO fake dataset ``observed spectrum" is produced. A fit is then done to that simulated dataset under both a normal hierarchy hypothesis and an inverted hierarchy hypothesis, with all neutrino mixing parameters allowed to vary as part of the fit according to Table~\ref{table:values}. In addition, a hypothesized $\sigma_{x}$ value is varied as part of the fit. The best-fit $\chi^{2}$ values for the normal-hierarchy fit and the inverted-hierarchy fit are then compared to produce a $\Delta \chi^{2}_{\mathrm{MH}}$ for finding the level of statistical separation JUNO will be able to achieve between the two hierarchies. This process is then repeated for each true $\sigma_{x}$ value many times to create multiple ``universes".
The distribution of results across these simulated universes characterizes JUNO's ability to measure
the mass hierachy and $\sigma_{x}$ accounting for both experimental uncertainty and uncertainty on the
mixing parameters.

\section{Results and Discussion} \label{results}
\begin{figure*}[t]
\centering
    \centering
     \includegraphics[width=.49\textwidth]{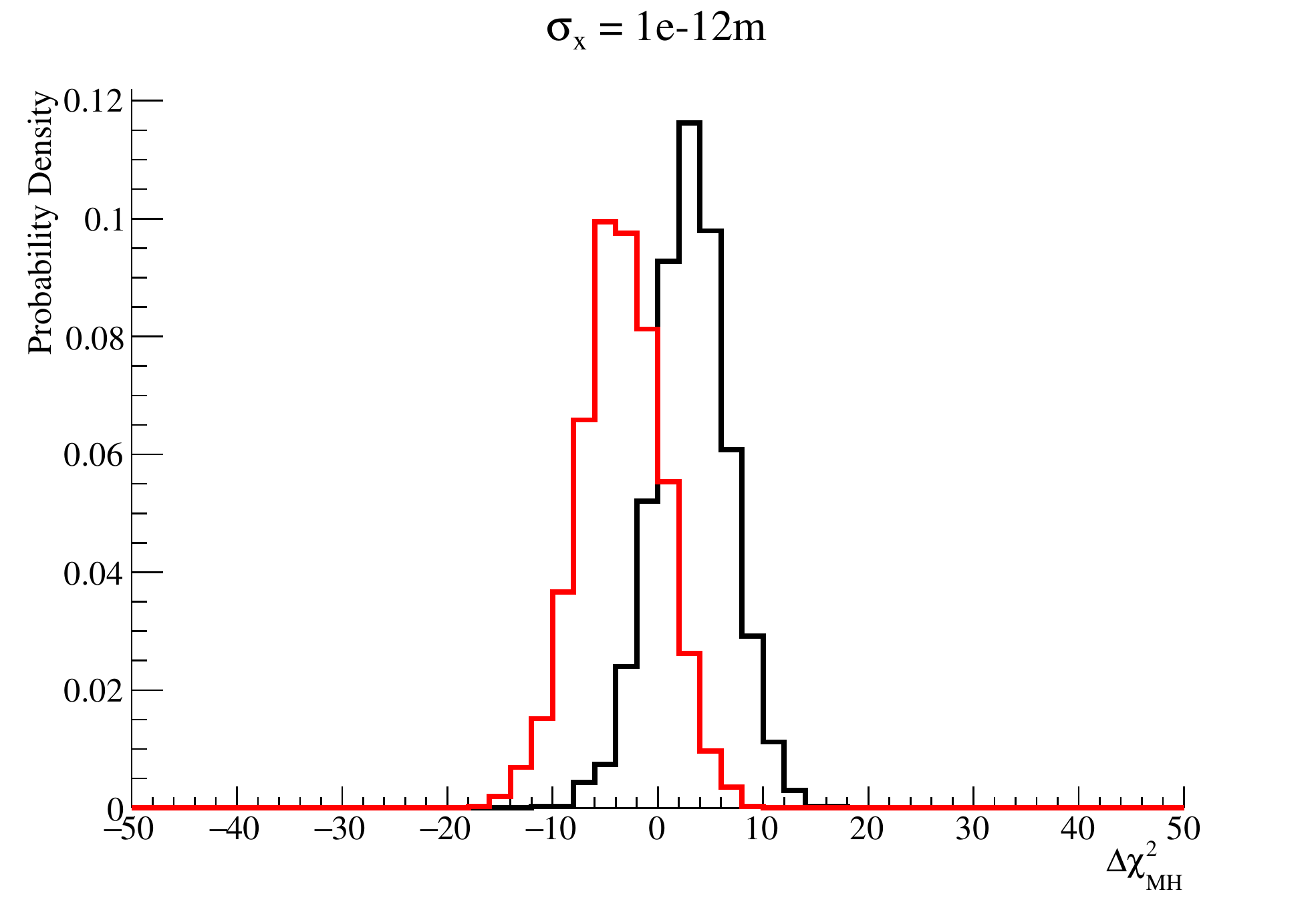}
     \includegraphics[width=.49\textwidth]{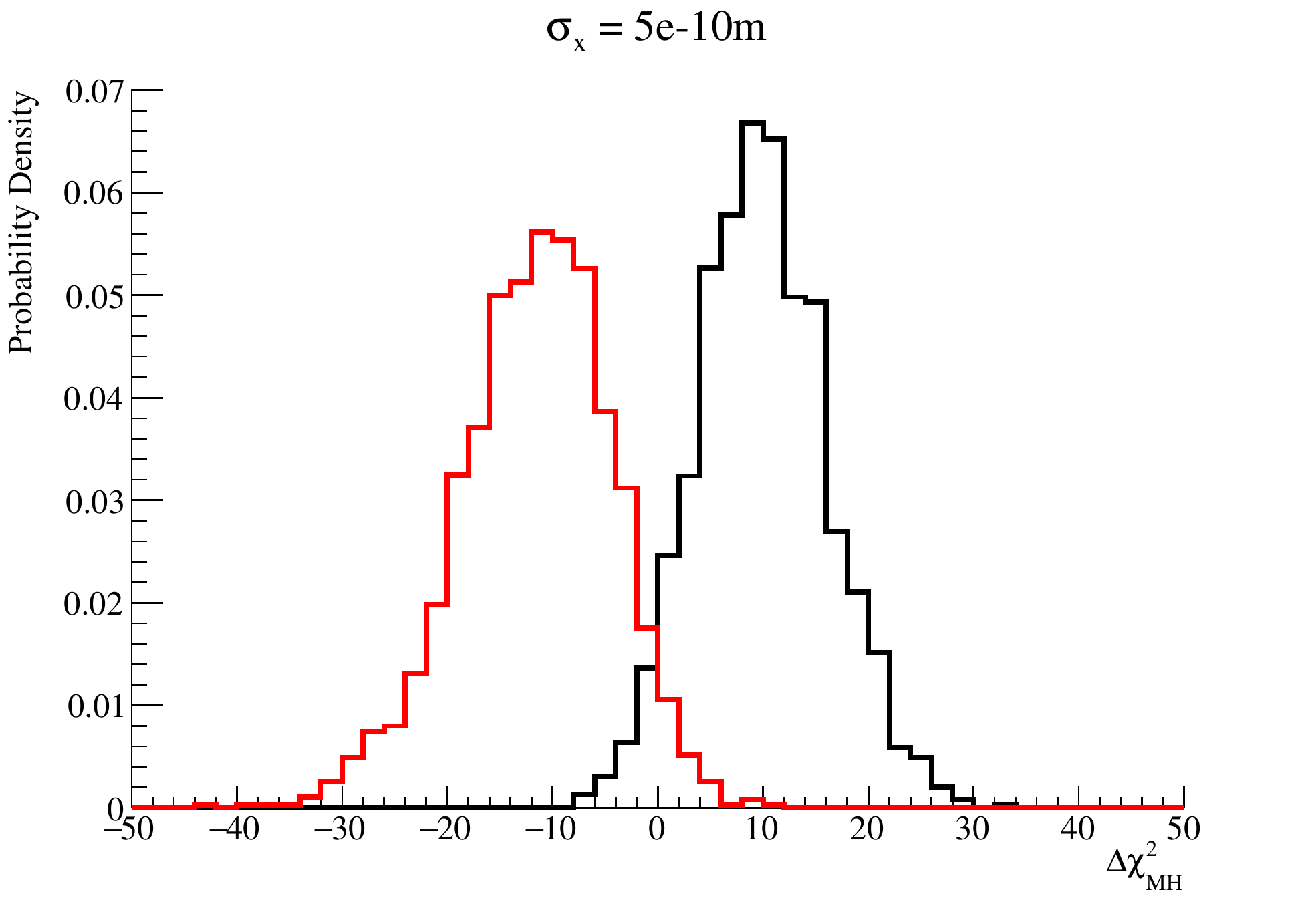}
     \label{figure_small_big_gaussians}
 \caption{Distribution of $\Delta\chi^{2}_{\mathrm{NH}}$ (black) and $\Delta\chi^{2}_{\mathrm{IH}}$ (red) for two different values of $\sigma_{x}$, $10^{-12}$\,m~(left) and $5\cdot10^{-10}$\,m~(right). For JUNO, $\sigma_{x} = 5\cdot10^{-10}$\,m is effectively the same as plane-wave oscillations.}
 \label{figure_example_gaussians}
\end{figure*}

\begin{figure*}[t]
\begin{centering}
\includegraphics[width=.49\textwidth]{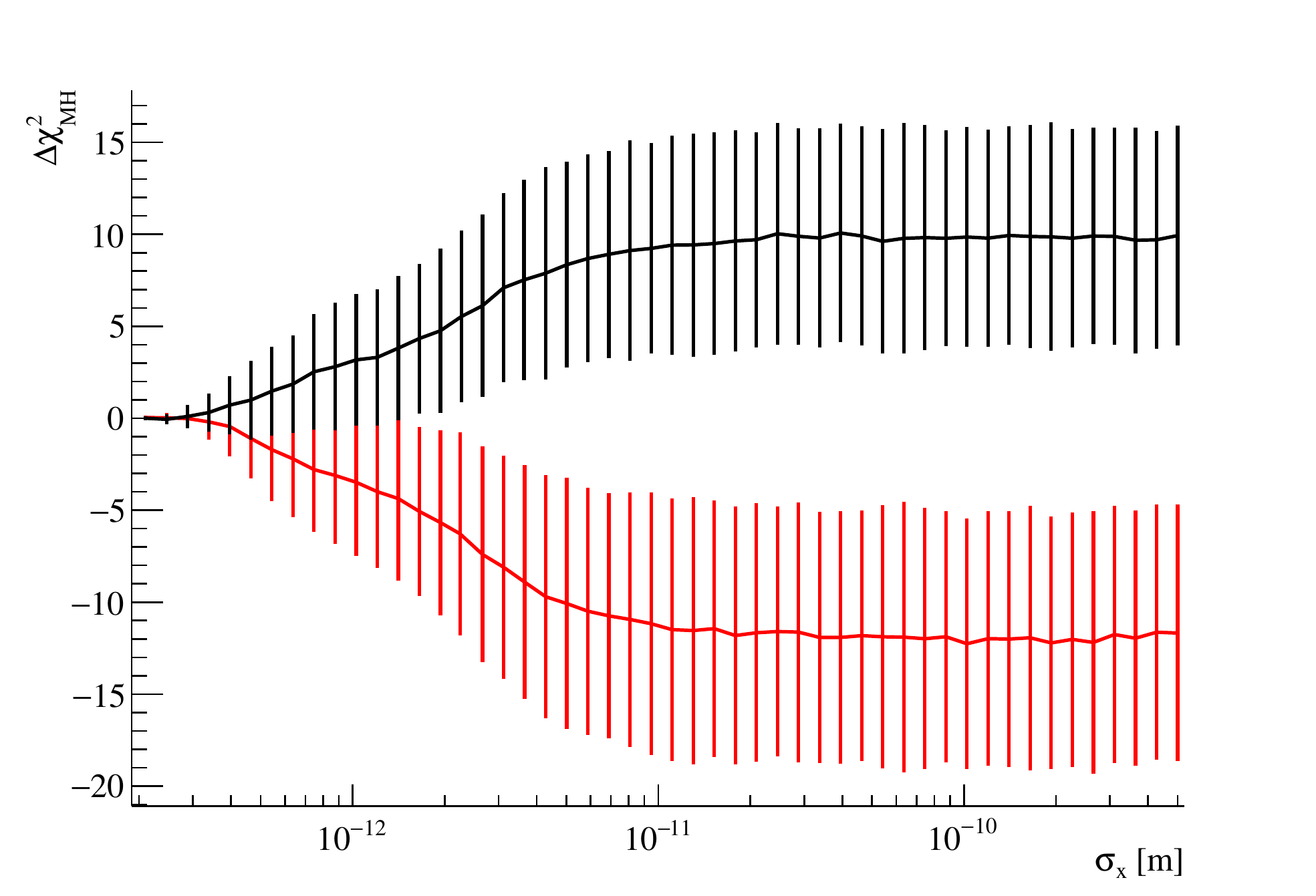}
\includegraphics[width=.49\textwidth]{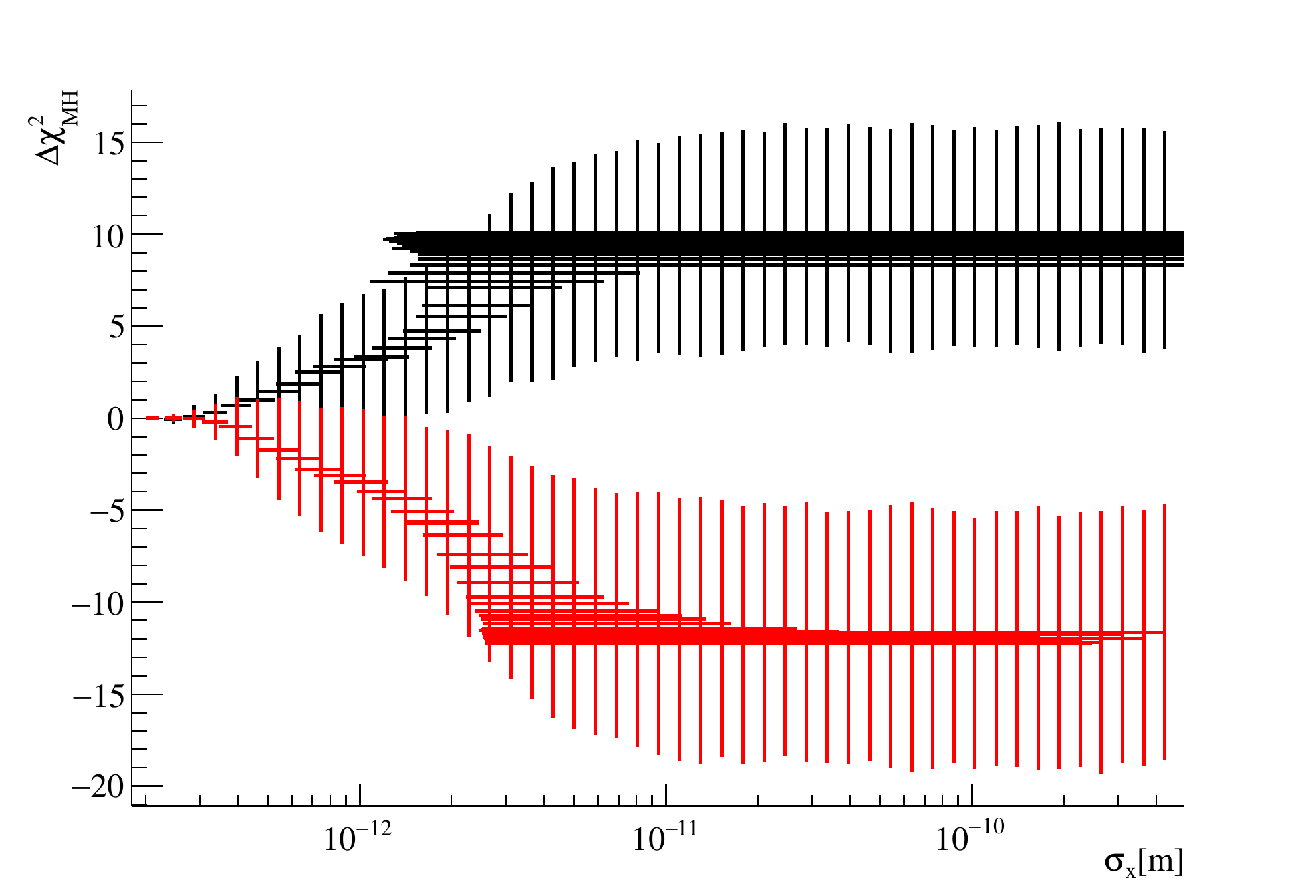}
\vspace{0cm}
\caption{(Left) JUNO's ability to determine the neutrino mass hierarchy as a function of true $\sigma_x$ for the case of an underlying normal hierarchy (black; $\Delta\chi^{2}_{\mathrm{NH}}$) and an underlying inverted hierarchy (red; $\Delta\chi^{2}_{\mathrm{IH}}$). (Right) The same plot, but now showing JUNO's simultaneous ability to measure both $\sigma_x$ and determine the hierarchy as a function of true $\sigma_x$ value. For $\sigma_x$ values above about $4\cdot 10^{-12}$~m, only a lower limit can be obtained. The horizontal error bars represent the bounds of the 68\% most central fit results.}

\label{figure_hierarchy}
\end{centering}
\end{figure*}

The $\Delta\chi^2_{\mathrm{NH}}$ and $\Delta\chi^2_{\mathrm{IH}}$ distributions of universes are shown with two example $\sigma_x$ values, $10^{-12}$\,m and $5\cdot10^{-10}$\,m, in Figure~\ref{figure_example_gaussians}. As can be seen, the distinction between the hierarchies is significantly degraded in the case of $\sigma_x=10^{-12}$\,m ($|<\Delta \chi^2_{\mathrm{NH}}>|\sim3.2$ and $|<\Delta \chi^2_{\mathrm{IH}}>|\sim3.4$). In contrast, for $\sigma_x=5\cdot10^{-10}$\,m, the wavepacket effect on oscillation probability is negligible, and therefore consistent with the plane-wave treatment, with $|<\Delta \chi^2_{\mathrm{NH}}>|\sim10$ and $|<\Delta \chi^2_{\mathrm{IH}}>|\sim12$. The results across a large range of $\sigma_x$ values are shown in Figure~\ref{figure_hierarchy} (left) in terms of both $\Delta\chi^2_{\mathrm{NH}}$ and $\Delta\chi^2_{\mathrm{IH}}$ as a function of true $\sigma_x$. The error bar on each point represents the RMS of the distribution of universes while the central value represents the mean.
For $\sigma_x>10^{-11}$\,m, the nominal hierarchy determination capability of JUNO is largely unaffected by the finite size of the wavepacket. However, the sensitivity to the hierarchy rapidly degrades for $\sigma_x<10^{-11}$\,m, until it disappears around $\sigma_x\sim10^{-12}$\,m.

\begin{figure}[h]
\begin{centering}
\includegraphics[width=9.3cm]{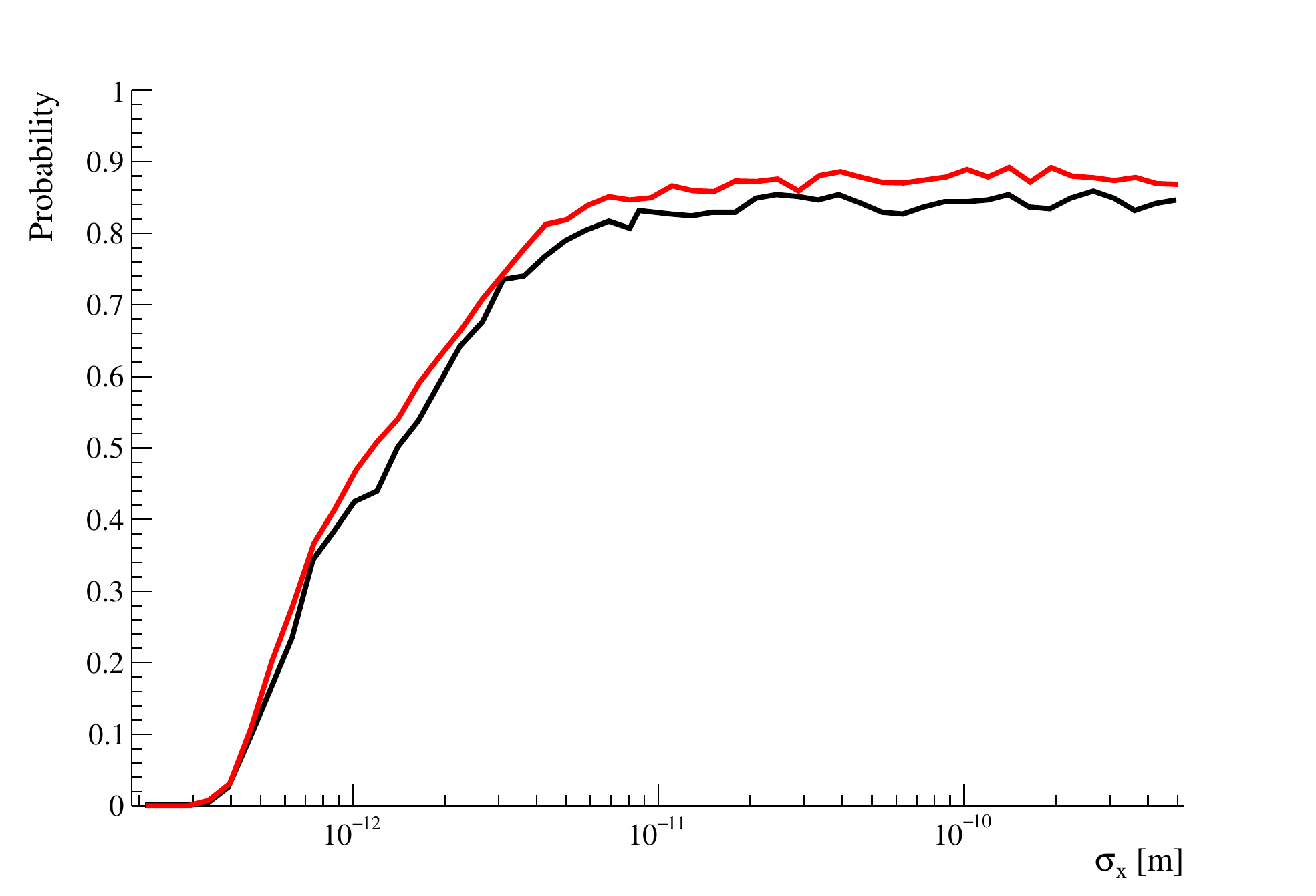}
\vspace{-.6cm}
\caption{The probability that JUNO will reject the incorrect hierarchy
with at least 95\% confidence as a function of $\sigma_{x}$ for a true normal hierarchy (black) and a true inverted hierarchy (red).}
\label{figure_ninefive_confidence}
\end{centering}
\end{figure}

\begin{figure}[t]
\begin{centering}
\includegraphics[width=9.3cm]{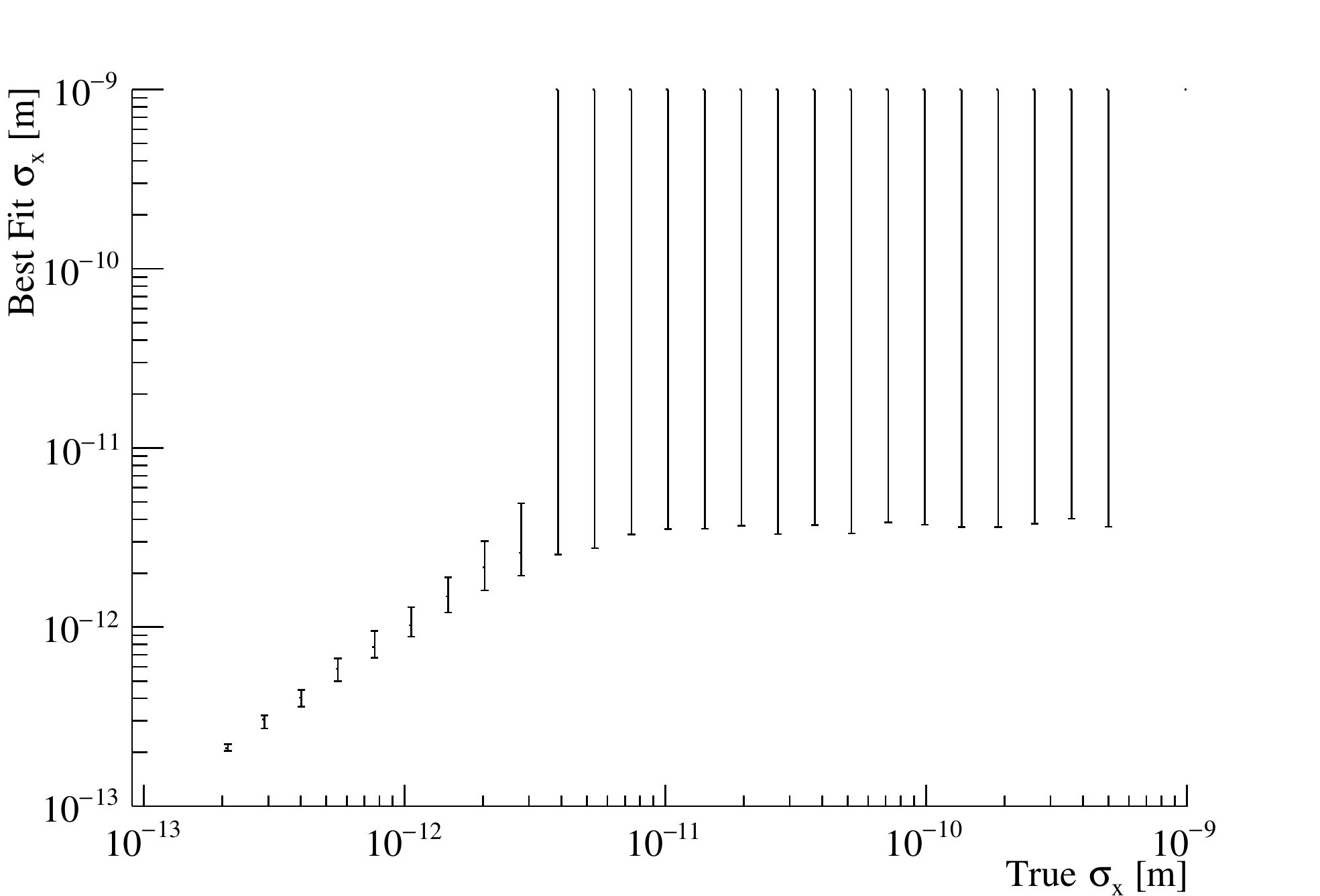}
\vspace{-.6cm}
\caption{The 1$\sigma$ measurement interval JUNO will be able to place on $\sigma_{x}$ as a function of true $\sigma_{x}$. 
For the points with error bars extending above $\sigma_{x}=1\cdot10^{-9}$~m, an achievable lower limit measurement is depicted.}
\label{figure_sigmax_resolution}
\end{centering}
\end{figure}

As might be expected from Figure~\ref{rate_plot}, there is a range of true $\sigma_x$ values for which JUNO is sensitive to both $\sigma_x$ and the orientation of the hierarchy at the same time. Below this range, JUNO is \text{only} sensitive to $\sigma_x$, and above this range, JUNO is \text{only} sensitive to the hierarchy orientation. Figure~\ref{figure_hierarchy} (right) again shows the behavior of  $\Delta\chi^2_{\mathrm{NH}}$ and $\Delta\chi^2_{\mathrm{IH}}$ as a function of true $\sigma_x$, but with the $\sigma_x$ measurement resolution achievable displayed as well. JUNO can produce a two-sided constraint on $\sigma_x$ for $\sigma_x<3\cdot10^{-12}$\,m. This result is reasonably consistent with Refs.~\cite{deGouvea:2020hfl,JUNO:2021ydg}. Figure~\ref{figure_ninefive_confidence} shows the probability that JUNO will be able to resolve the neutrino mass hierarchy with 95\% confidence at different
$\sigma_{x}$ values for the cases of a true and inverted hierarchy. As can be seen, wavepacket effects detract from JUNO's hierarchy determination capability as far as nearly two orders of magnitude above the current experimental lower limit, up to about $\sigma_x=10^{-11}$\,m. Fig.~\ref{figure_sigmax_resolution} shows the 1$\sigma$ confidence
interval on $\sigma_{x}$ that JUNO is likely to produce at different
$\sigma_{x}$ values. At $\sigma_x=5\cdot10^{-13}$\,m, for example, the expected achievable measurement resolution is $\delta(\sigma_x)=1.7\cdot10^{-13}$\,m. These figures show that for $\sigma_{x}$ values between $1\cdot10^{-12}$\,m and $3\cdot 10^{-12}$\,m JUNO will likely be able to resolve the neutrino mass hierarchy at 95\% confidence and produce a measurement of $\sigma_{x}$.

While sensitivity to the hierarchy can be reduced by finite $\sigma_x$, JUNO's expected leading measurements of both $\sin^2\theta_{12}$ and $\Delta m^2_{21}$, represented by the depth and energy of the IBD event rate dip around $E_{\overline{\nu}_e}=3$\,MeV in Figure~\ref{rate_plot}, respectively, are not significantly changed for any experimentally allowed value of the parameter ($\sigma_x>2\cdot10^{-13}$\,m)~\cite{deGouvea:2021uvg} since, as previously discussed (see Eq.~\ref{simple_osc}), the slower, solar oscillations, with frequency governed by $\Delta m^2_{21}$, are less affected by $\sigma_{x}$. 

\begin{figure}[h]
\begin{centering}
\includegraphics[width=9.3cm]{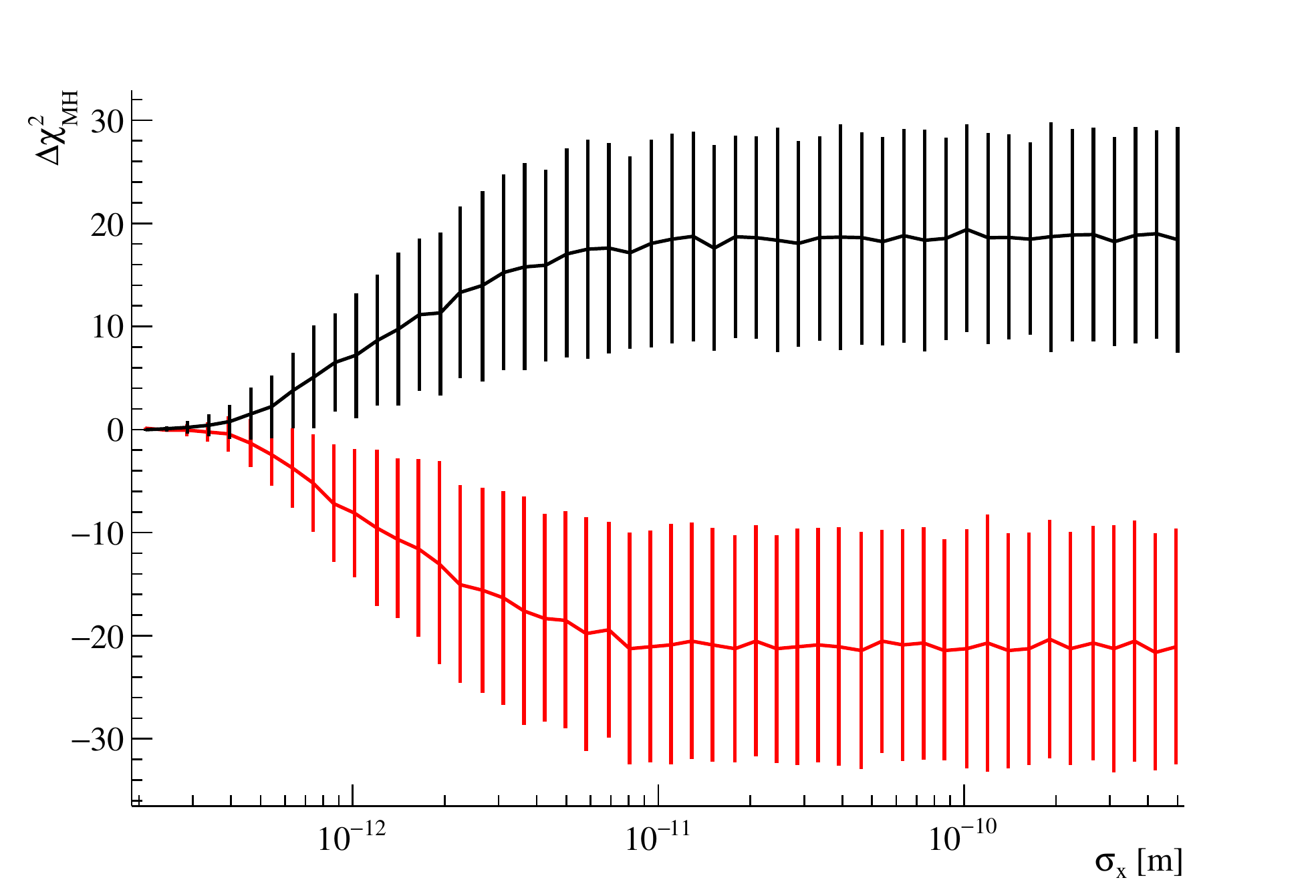}
\vspace{-.6cm}
\caption{In the case that the worldwide precision on $\Delta m^2_{3l}$ improves by a factor of two [$\delta(\Delta m^2_{3l})=(0.028\rightarrow0.014)\cdot 10^{-3}~\mathrm{eV}^2$], JUNO's ability to determine the hierarchy as a function of true $\sigma_x$ value, for both the case of an underlying normal hierarchy (black) and an underlying inverted hierarchy (red).}
\label{figure_better_precision}
\end{centering}
\end{figure}

Similar to $\Delta m^2_{3l}$, a measurement of $\theta_{13}$ in JUNO, represented by the amplitude of the small, atmospheric oscillations in Figure~\ref{rate_plot}, would be severely affected by $\sigma_x$. Fortunately, $\theta_{13}$ has been precisely reported by the Daya Bay experiment~\cite{DayaBay:2018yms} and, since JUNO cannot expect to improve upon this measurement, even in the absence of wavepacket effects~\cite{JUNO:2015zny,JUNO_2022_physics_paper}, it will rely on this external result. Notably, however, JUNO's sensitivity to both $\sigma_x$ and the hierarchy, considered individually or together, can be greatly improved with additional, future constraints on $\Delta m^2_{3l}$ from long-baseline accelerator and atmospheric experiments~\cite{T2K:2011qtm,NOvA:2021nfi,DUNE:2020jqi,Super-Kamiokande:2019gzr,Hyper-Kamiokande:2018ofw,IceCube:2016zyt}. These experiments, which primarily rely on pion decay-in-flight neutrinos traveling 100s-10000s of km, are insensitive to the wavepacket effect, in the sense that it won't alter their oscillation parameter measurements~\cite{Jones:2014sfa}. Figure~\ref{figure_better_precision} again shows $\Delta\chi^2_{\mathrm{NH}}$ and $\Delta\chi^2_{\mathrm{IH}}$ as a function of true $\sigma_x$, but with a 50\% reduction in the uncertainty on $\Delta m^2_{3l}$ compared to the current value(s) shown in Table~\ref{table:values} [$\delta(\Delta m^2_{3l})=(0.028\rightarrow0.014)\cdot 10^{-3}~\mathrm{eV}^2$], consistent with projections from a low-exposure DUNE measurement (300\,kton$\cdot$MW$\cdot$years)~\cite{DUNE:2020jqi}, at least, but also roughly representative of expected near-term global improvements informed by T2K~\cite{T2K:2011qtm}, NOvA~\cite{NOvA:2021nfi}, and IceCube~\cite{IceCube:2016zyt}.

Of course, almost any new neutrino physics relevant at these baselines and energies, such as (e.g.) non-standard neutrino interactions, neutrino decay, and/or oscillations involving a sterile flavor, seriously encumbers the use of long-baseline $\nu_\mu \rightarrow \nu_\mu$ and $\nu_\mu \rightarrow \nu_e$ results as inputs for performing medium-baseline reactor-based measurements of $\overline{\nu}_e \rightarrow \overline{\nu}_e$. Indeed, any deviation to the three-neutrino paradigm, or standard neutrino interaction paradigm insofar as it is used to extract oscillation measurements, will lead us to revisit many existing results and future sensitivities. While a $\sigma_x$ measurement and/or hierarchy determination may be confused in such a situation, JUNO's plethora of high precision results will certainly form an essential global input for either probing this new physics or overly constraining the three neutrino oscillation paradigm.

Aside from improved external measurements of $\Delta m^2_{3l}$ (and/or the orientation of the hierarchy itself), realistic future sensitivity enhancements to JUNO in terms of $\sigma_x$, the hierarchy, or $\sigma_x$ and the hierarchy considered simultaneously, other than ``more data" given the statistics-limited measurement, are somewhat difficult to envision. One could perhaps consider a smaller spread in reactor baselines, far-detector upgrades, and/or a reliable theoretical prediction for $\sigma_x$, as potential sources of sensitivity improvements. The spread in source-to-detector distances leads to oscillated spectrum smearing, with the largest bin migration coming from the currently operating 17.4~GWth Daya Bay reactor complex at 215~km (6.4\% of the expected IBD events~\cite{JUNO_2022_physics_paper}). Although it is hard to imagine that this complex stops producing power for any significant length of time, which would enhance JUNO's sensitivity, perhaps the JUNO experiment will start taking data before the other long(er)-baseline complex at Huizhou (17.4~GWth at 265~km) comes online in 2025. Improvements to the JUNO energy resolution, primarily driven by photon statistics, would strengthen sensitivity, but JUNO is already at the cutting edge of what is possible with a multi-kiloton-scale detector (77.9\% photocoverage, with 17612 20'' PMTs @ 34\% quantum efficiency~\cite{JUNO:2015zny,JUNO_2022_physics_paper}). A reliable theoretical prediction for $\sigma_x$ could also offer improved experimental sensitivity to the value and the orientation of the hierarchy, but only in and near the ``transition region" presented above ($\sim10^{-12}<\sigma_x<10^{-11}$\,m), in which the experiment has some sensitivity to both. Even in the case that $\sigma_x$ is perfectly predicted, sensitivity to the hierarchy cannot be achieved by JUNO significantly below this range. 

In terms of JUNO ``upgrades" towards improved sensitivity, it is worth emphasizing the need for a near detector, which is assumed for this analysis, especially in consideration of the possible reactor flux substructure and uncertainties associated with the $5<E_{\bar{\nu}_e}<7$~MeV ``reactor bump"~\cite{prospect2,stereo,NEOS:2016wee,reno,dayabay,doublechooz}. Measurements from the near detector JUNO-TAO (2.8\,tons, 30\,m from a 4.6\,GWth reactor core of the Taishan Nuclear Power Plant; 2000~IBD events per day; sub-1\% energy resolution) are expected to constrain the reactor flux bin-to-bin shape uncertainties at the sub-1\% level and thereby improve hierarchy sensitivity by $\Delta \chi^2\sim1.5$ compared to without~\cite{JUNO:2020ijm}. In the absence of this near detector, the reactor flux shape uncertainties are expected at the 2\% bin-to-bin level using the Daya Bay reference spectrum, noting Daya Bay's energy resolution of $0.08/\sqrt{E_{\mathrm{vis}}~\mathrm{(MeV)}}$~\cite{JUNO:2020ijm}. However, even with a precision near detector (proximal to a \textit{single} core), uncertainties in the bin-to-bin migration due to the different fission fractions among the many contributing reactor cores, and associated time dependence of these fractions, may be significant. 
Indeed, reactor flux predictions and their uncertainties are still evolving substantially in time (for a recent example, see Ref.~\cite{Kopeikin:2021ugh}), with the reliant relevant oscillation parameter sensitivities and measurements, including in short- and medium-baseline experiments, following suit (see, e.g., Refs.~\cite{Berryman:2020agd,Giunti:2021kab}). In general, these issues point to the continued need for theoretical and experimental work towards understanding reactor antineutrino production.

\section{Conclusion}
We have presented the JUNO sensitivity to both the orientation of the neutrino mass hierarchy in the presence of wavepacket effects and the size of the wavepacket itself. We find that wavepacket effects will detract from the hierarchy determination up to nearly two orders of magnitude above the current experimental lower limit, and that JUNO will be highly capable of precisely measuring the size of the wavepacket in large regions of currently allowed parameter space. In addition, external experimental measurements, in particular from long-baseline oscillations, will significantly enhance these physics capabilities of JUNO. Given the demonstrated importance of this non-exotic, ``standard" quantum mechanics effect on JUNO, and the experiment's expected contributions to the worldwide neutrino oscillation program, without even mentioning the relevance of the hierarchy determination to cosmological- and terrestrial-based probes of neutrino mass, we would like to encourage more study of the electron antineutrino wavepacket, in particular towards forming a reliable theoretical prediction of its characteristic size at creation inside a nuclear reactor.

    

\section{Acknowledgements}
We thank C.A.~Arg{\"u}elles and B.J.P. Jones for useful discussions. This work is supported by the Department of Energy, Office of Science, under Award No. DE-SC0007859. 

\bibliography{main}

\end{document}